\documentclass[prl,twocolumn,showpacs]{revtex4}
\usepackage{graphicx}
\usepackage{amsmath}

\begin{document}

\title{Novel exponents control the  quasi-deterministic limit of the extinction transition}

\author{David A. Kessler}
\affiliation{Department of Physics, Bar-Ilan University, Ramat-Gan IL52900, ISRAEL}
\author{Nadav M. Shnerb}
\affiliation{Department of Physics, Bar-Ilan University, Ramat-Gan IL52900, ISRAEL}

\begin{abstract}
The quasi-deterministic limit of the generic extinction transition
is considered within the framework of standard epidemiological
models. The susceptible-infected-susceptible (SIS) model is known to exhibit
a transition from extinction to spreading, as the infectivity is increased, described by the
directed percolation equivalence class.  We find that  the distance from the transition point, and the prefactor controlling the divergence of the (perpendicular) correlation length, both
scale with the local population size, $N$, with two novel universal exponents. Different
exponents characterize the large $N$ behavior of the
susceptible-infected-recovered (SIR) model, which belongs to the
dynamic percolation class. Extensive numerical studies in a range of
systems lead to the conjecture
that these characteristics are generic and may be used in order to
classify the high density limit of any stochastic process on the
edge of extinction.
\end{abstract}
\maketitle

\section {Introduction}
Perhaps the most useful approximation in science is the description
of microscopic, stochastic processes via "mean field" deterministic
equations. Except for the most fundamental parts of particle
physics, almost any other study involves that procedure: classical
mechanics is used to describe quantum systems, classical
electromagnetism neglects the fluctuations of the photon density and
continuum mechanics averages out the microscopic stochasticity
involved in the motion of an individual molecule. Chemical reaction
kinetics is generally described by rate equations; population
dynamics and other ecological processes are depicted using the
concept of logistic growth, or by Lotka-Volterra equations. The
dynamics of the individual, microscopic, constituents is always
stochastic, and is subject to fluctuations; these fluctuations are
smeared out when the system is described by deterministic rate
equations.

Basically, the underlying assumption beyond all these approximations
is that the microscopic fluctuations are averaged out in the ``large
density'' (many atoms, animals, quanta) limit. A generic analysis of
the deterministic limit, like a $1/N$ expansion where $N$ is the
number of elements, is still lacking in many fields. Problems like
the quantum classical correspondence in chaotic systems, or the
decay of a quasistationary state to an absorbing state are still a
subject of intensive studies. The situation becomes even more
complicated when spatially extended systems, made of diffusively
coupled patches, are considered. What exactly determine the ``large
N'' limit in that system? Should the number of microscopic entities
be large on a single patch, or within a correlation length? All
these questions are still open, as no systematic perturbation theory
in $1/N$ exists so far.

In this paper we consider a generic process in population dynamics:
the extinction transition with an absorbing state. The framework
used is two well-known models for epidemics, namely, the SIS and the
SIR models \cite{sa,am}. For a well mixed population of size $N$
(say, on a single patch) the stochastic process starts by the
introduction of a single infected (``I'') individual into the system.
This individual may infect any other, susceptible (``S''), individual
with rate $\alpha/N$, where after the infection the susceptible
becomes infected and may infect other susceptible members of the
community; the only other process is a "recovery" of an infected
person; this happens with rate $\beta$. In the SIR model, the
recovered individuals are then immune against the disease, while the
SIS model describes the case where the recovered became susceptible
again. The elementary processes, thus, are
\begin{eqnarray} \label{eq1}
 S+I \stackrel{\alpha/N}{\rightarrow} 2I  \qquad I \stackrel{\beta}{\rightarrow} \emptyset \qquad (SIR) \nonumber \\
  S+I \stackrel{\alpha/N}{\rightarrow} 2I  \qquad I
\stackrel{\beta}{\rightarrow} S \qquad (SIS)
 \end{eqnarray}
Denoting by $I$ the fraction of infected individuals, and by $S$ the
fraction of susceptibles,  the mean field equations in the well mixed
limit are
\begin{eqnarray} \label{eq2}
\frac{dI}{dt} = \alpha I S - \beta I  &\qquad& \frac{dS}{dt} = -\alpha I S + \beta I  \quad  (SIS) \nonumber \\
\frac{dI}{dt} = \alpha I S - \beta I &\qquad& \frac{dS}{dt} =
-\alpha I S \quad  (SIR).
 \end{eqnarray}
Clearly, the SIR process is self-limiting, as $S$ decreases with
time, while the SIS process may support an endemic state with $I =
1- \alpha / \beta$. Another piece of information garnered from the
rate equations is the existence of a transition when the infectivity parameter  $R_0 \equiv
\alpha  / \beta$ crosses the value $R_c = 1$. Below $R_0=R_c$, both SIS and SIR processes immediately
decay; above that value, an outbreak is possible.

The mean field equations (\ref{eq2})  are an approximation for the
real stochastic process (\ref{eq1}). Some characteristics of the
epidemic, though,  may be calculated exactly for the real stochastic
process in the well mixed limit. In particular,

$ \bullet$ Below $R_c$ both
processes are the same in the large $N$ limit, since the change of a susceptible becoming reinfected in the SIS model is vanishingly small.

 $ \bullet$ For large $N$, above $R_c$ there are two peaks in the distribution of epidemic sizes.  The first peak is at size 1 (i.e. the initial infected individual, only).  The second peak for the
 SIR model is at the size predicted by the deterministic equations~\cite{watson}.  For the SIS model,
 the second peak is at an exponentially large value, corresponding to the exponentially
 long lifetime of the metastable endemic state\cite{mysis}.

$ \bullet$ At $R_c$, the average size of the epidemic scales for large $N$ as $N^{1/2}$
for the SIS model~\cite{dlsis,mysis} and as $N^{1/3}$ for the SIR model~\cite{lof,bn,mysir}.

 Let us consider now  a one dimensional array of L patches, with $N$
 susceptible individuals on each patch, where a single infected individual is
 introduced at a single patch.

Above $R_c$, and for $N \to \infty$,  the deterministic equations
(\ref{eq2}), with an appropriate contact term playing the role of
 "diffusion" , give rise to a constant velocity solution.
 The mean field dynamics is akin to that  of the FKPP equation \cite{fisher,kpp}, and yield a
 velocity proportional to the square root of the distance from
 $R_c$.   The solution for the SIS case is that of a front separating the
 metastable uninfected state ahead of the front from the endemic state behind.  In the SIR case,
 the solution is a pulse, leaving behind a state with a diminished susceptible
 population such that the effective infectivity parameter, $R_0 S/N$, is below $R_c$.

What happens in the stochastic models then?  Let us discuss the SIS
case first. For $N=1$ the SIS model is equivalent to the contact
process~\cite{Harris}, where there is a transition to a propagating state at a
finite value of $R_0 > 1$.  This transition is known to be in the
directed peculation (DP) equivalence class~\cite{Hinr}.  Above the transition, there
is a finite chance of generating a wave of infection whose lifetime
is infinite.  The average velocity of the wave is less than that
predicted by the deterministic equation.  This has
been analyzed by Brunet and Derrida \cite{bd}, who show that as
$N\to \infty$, the average velocity approaches the deterministic value, albeit with
anomalously large ${\cal{O}}(\ln^{-2}N)$ corrections. Below the
directed percolation transition, the typical spatial extent of the
epidemic, denoted $\xi_\perp$, is finite, and diverges as the
transition point is approached.  Following the Janssen-Grassberger
conjecture \cite{jg}, it is widely  believed that \emph{any} generic
extinction transition, and, in particular, the transition for the
SIS model for \emph{any} N,  falls in the equivalence class of
directed percolation.

In order to test these predictions in the large density limit we
have simulated the SIS process on a one dimensional array of sites.
On each site there are  $N$ individuals, and this number is fixed
throughout the process. The epidemics is ignited by the introduction of a
single infected individual on a single site. The chance of an $I$ to
infect any susceptible individual on the same site is $(1-\chi)
\alpha/N$, and its chance to infect an $S$ on one of the two
neighboring sites is $\chi \alpha/N$. Thus,  $\chi$ is the
inter-site infectivity rate, while the population within a site is
considered as well mixed. The  chance of recovery  is independent of
the spatial structure and the recovery rate is $\beta$ for any
individual. We have used an exact, agent-based simulation in order
to study the approach to transition from below, using the divergence
of the perpendicular correlation length $\xi_\perp$ as a marker
of the transition.

What is the effect of increasing $N$ on the transition? Fixing
$\beta$ and $\chi$ one should expect a dependence of  $R_{DP}$, the value of $R_0$ at
the DP transition,
 on $N$. If $N=1$, for example,
the SIS becomes a simple contact process on a line, and the DP
transition happens when the infection rate is 3.297 times higher than
the recovery rate; accordingly, $R_{DP}(N=1)  =
3.297/\chi$. On the other hand, as $N \to \infty$, $R_{DP}$ should
approach the value $R_c \equiv 1$ and becomes  $\chi$ independent,
as a single site may support the endemic state. $R_{DP}$ should
interpolate between these two limits as a function of $N$.

Another effect has to do with the divergence of $\xi_\perp$ at the
transition. According to the stochastic theory, right below the
transition $\xi_\perp$ is finite but large. On the other hand, the
classical equations predict that $\xi_\perp = 0 $ below the
transition. One expects both statements to be true at the large $N$
limit. The only way out of that paradox is to understand that the
classical description must fail close enough to the transition, but
the region in which  it fails must shrink to zero in the $N \to
\infty$ limit. Say it another way, at a fixed distance from the
transition, and for large enough $N$, the perpendicular correlation
length should approach zero.

\begin{figure}
\center{\includegraphics[width=0.45\textwidth]{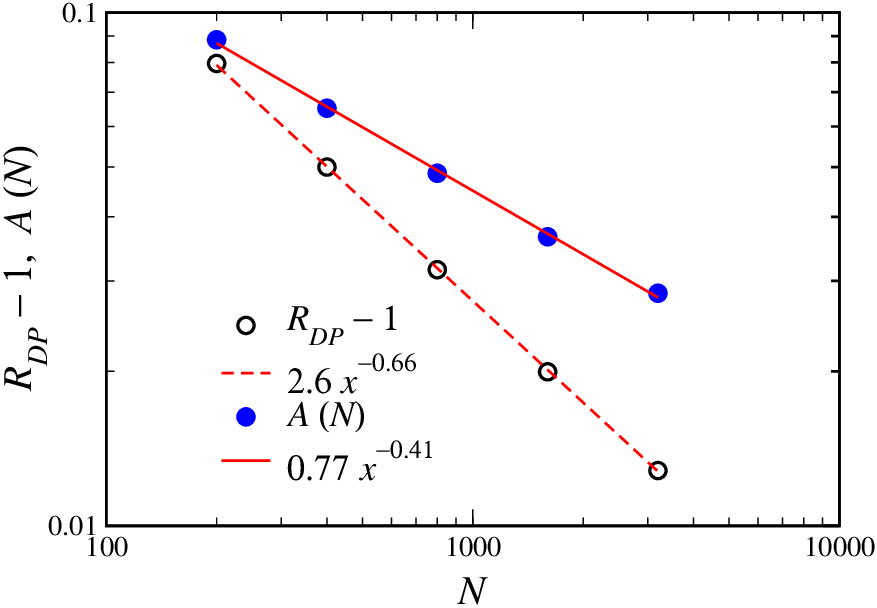}}
\caption{Directed percolation transition point, $R_{DP}$, of the 1D SIS model as a function of $N$, together with the critical perpendicular correlation length amplitude $A(N)$.  Also shown are power-law fits to the data. $\xi=0.2$}
\label{rdp}
\end{figure}

Our simulations support all parts of this picture. First, the
Jansen-Grassberger conjecture works and in all cases studied  we
find that $\xi_\perp$ diverges as the transition point is approached
from below with the expected power $-\nu_\perp\approx -1.097$.
Second, we find that  $\xi_\perp$  obeys a universal scaling with
$N$, independent of the details of the model. Specifically, near the
transition
\begin{equation}
\xi_\perp = A(N) (R_{DP}(N) - R_0)^{-\nu_\perp}
\label{trans}
\end{equation}
where for large $N$
\begin{equation}
R_{DP} (N) - R_c\sim  N^{-\kappa}, \qquad \kappa \approx 0.66
\end{equation}
where for large $N$
\begin{equation}
A(N) \sim N^{-\tau}, \qquad \tau \approx 0.41.
\end{equation}
The data supporting these findings are presented in Fig. \ref{rdp} for the
specific case of  an SIS model with fixed recovery time and $\chi =
0.2$. The applicability of these results, though, is much wider; the
same exponents turn out to describe the large $N$ limit of the SIS
model with different $\chi$ and with exponentially distributed
recovery time, as well as other stochastic  models like
branching-annihilation-death ($A \to \emptyset, \  A+A \to
\emptyset,  \  A \to 2A$) and so on. We thus conjecture that these
exponents describe the large $N$ behavior of \emph{any}
one-dimensional extinction transition that belongs to the DP
equivalence class.

\begin{figure}
\center{\includegraphics[width=0.45\textwidth]{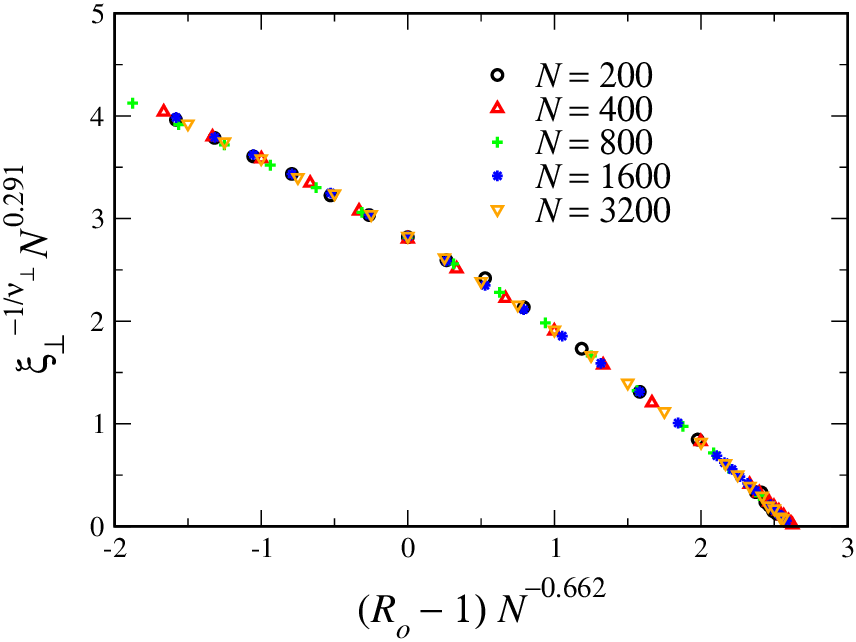}}
\caption{Collapse of the 1D SIS model scaled correlation length, $\xi_\perp^{-1/\nu_\perp} N^{0.284}$
versus the scaled infectivity $(R_o-1)N^{0.662}$ for $N=200$, $400$, $800$, $1600$ and $3200$. The contact parameter is again $\chi =0.2$, and the fixed recovery time variant was used. 
}
\label{sisxicol}
\end{figure}

The data collapse presented in Figure \ref{sisxicol} reveals the
existence of much stronger regularity at large $N$'s. Eq. (\ref{trans})
implies that $N^{\kappa-\tau/\nu_\perp}\xi_\perp^{-1/\nu_\perp}$ is
a linear function of $(R_o - R_c)N^\kappa$ close to the transition. In
fact, this result generalizes to a whole scaling regime where the
scaled correlation length  is a function of  the scaled distance
from the classical transition:
\begin{equation}
N^{\kappa-\tau/\nu_\perp}\xi_\perp^{-1/\nu_\perp} = F( (R_o - R_c)N^\kappa)
\end{equation}
Again, we have verified that this scaling behavior is independent of
the (nonzero) strength of the intersite coupling.

\begin{figure}
\center{\includegraphics[width=0.45\textwidth]{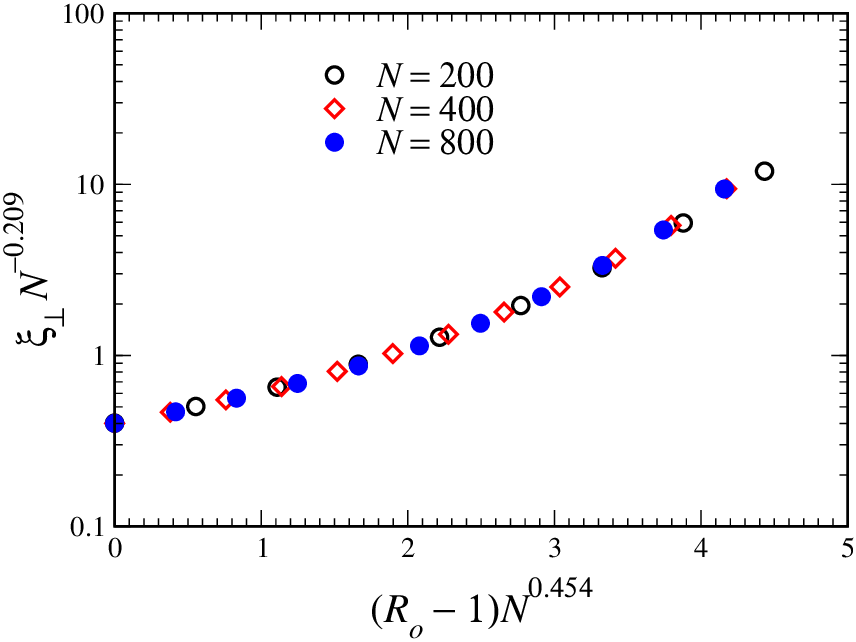}}
\caption{Collapse of the SIR scaled correlation length, $\xi_\perp N^{-0.209}$
versus the scaled infectivity $(R_o-1)N^{0.454}$ for $N=200$, $400$, and $800$. Here too  $\chi =0.2$.}
\label{sirxicol}
\end{figure}

We now turn to examine the behavior of the stochastic $SIR$ model at
large $N$.  Here, the model is supposed to be in the dynamic
percolation universality class~\cite{gras83,grascar85}.  This class does not have a
transition in one spatial dimension, and the propagating pulse
always dies out in an infinite system.  Still, the correlation
length exhibits a scaling behavior with $N$ very similar to that of
the SIS model.  We find that the scaled correlation length,
$\xi_\perp N^{-0.209}$ is a function of the scaled infectivity $(R_o
- R_c)N^{0.454}$.  The data collapse is presented in Fig.
\ref{sirxicol}. Here the correlation length is finite for all
infectivities, indicating that we are indeed below the percolation
transition.  The correlation length does however grow very rapidly with $R_o$.

In the simulations presented above, we have used as our measure of the correlation length the inverse of the
exponential falloff rate of the distribution function for infection location.  It is remarkable
that in both models, for all $R_o$ studied (below the extinction transition), this distribution function appears to be given {\em{exactly}} (at least to within our
statistical errors, over some seven decades) by the functional form
$P(x) = C(R_o,N) e^{-|x-x_o|/\xi_\perp(R_o,N)}$, with $x_o$ is the location of the original infected person.  The average {\em total} number of infections is given then roughly by ${\cal{N}}=2C(R_o,N)\xi_\perp(R_o,N)$ (assuming
$\xi_\perp\gg 1$, as it is for $N$ large).  We can then deduce the scaling properties of $C(R_o=1,N)$ with $N$.  The number of infections at the central site, $C(1,N)$ are the result of the on-site infections nucleated by the original infected person and by those nucleated by the neighboring sites.  The effective on-site infection rate is $R_o(1-\chi)$,
and since this is below $R_c=1$, each initial infection develops essentially independently, giving rise to $1/(1-R_o(1-\chi))$ total infections.  The number of primary infections induced by the neighbors is $\chi C \exp(-1/\xi_\perp) \approx \chi C (1 - 1/\xi_\perp)$.  Thus, we have, for $R_o=1$
\begin{eqnarray}
C &=& \frac{1}{1-(1-\chi)} \left(1 + \chi C \left(1 - \frac{1}{\xi_\perp}\right)\right)\\
&=& C + \frac{1}{\chi} - \frac{C}{\xi_\perp}
\end{eqnarray}
This implies that
\begin{equation}
C = \frac{\xi_\perp}{\chi}
\end{equation}
so $C(1,N)$ scales exactly the same with $N$ as $\xi_\perp(R_o=1)$.  Thus, the total number of infections scales as $\xi_\perp^2(R_o=1)$, in both the SIS and SIR models.
For the SIS model, this gives the total number of infections scaling as $N^{2(\kappa\nu_\perp - \tau)} \approx N^{0.623}$, whereas in the SIR model, we have the total number of infections scaling as approximately $N^{0.418}$.  Notice that in both cases, the total
number of infections at the critical infectivity scales faster than the 0 dimensional results, $1/2$ and $1/3$, respectively.  We have tested these predictions in our simulations, (data not shown) and found them to be very well satisfied.  Also, preliminary work indicates that these exponents are higher still in two dimensions, and saturate at 1 as the
dimension goes to infinity~\cite{long}.

In the SIS model, as in the related contact process, the total number of infections diverges at the directed percolation transition, with the exponent $\gamma=\nu_\parallel + \nu_\perp - 2\beta=2.278$.  However, the total number of infections is equal to $C\xi_\perp$.  Thus, $C$ should diverge at the transition as $\gamma-\nu_\perp = 1.181$.  Combining this with our above result for the scaling of $C$ with $N$ at criticality, we get
that
\begin{equation}
(C N^{-0.311})^{-1/1.181} = {\cal{G}}\left((R-1)N^{0.668}\right)
\end{equation}
where ${\cal{G}}$ vanishes linearly at the transition.  It is remarkable that even though the directed percolation exponent associated with $C$ differs from that of $\xi_\perp$, the $N$ scaling is the same.

Our simulations suggest that the quasi-deterministic  region is
controlled by the new critical exponents $\kappa$ and $\tau$. At the
transition, the  deterministic limit does not exist and the
stochastic dynamics of the microscopic constituents determines the
system behavior for any time scale. Off transition, on the other
hand, the effect of stochasticity vanishes for large enough $N$. The
value of $\kappa$ and $\tau$ is parameter independent, as long as
the dimensionality of the system and the type of stochastic
transition (DP vs. dynamic percolation)  are kept fixed.

It is interesting to point out an exception to our analysis, namely,
the branching-annihilation process with no bare death term ($A
\stackrel{\sigma}{\rightarrow} 2A, A+A
\stackrel{\lambda}{\rightarrow} \emptyset$) for Brownian particles.
Here at low densities there is a DP type 
transition at finite birth rate, but as $N \to \infty$ (i.e.,
$\lambda \to 0$) the transition approaches $\sigma = 0$. The
deterministic limit of the transition in that case does \emph{not}
belong to the directed percolation equivalence class; in fact, it is
known that for $\sigma = 0$ the density decays like $t^{-1/2}$ in
one spatial dimension. Indeed, Cardy and Tauber~\cite{cardy} calculated that the
distance to the DP transition should scale as $N^{-2}$, (since close to the classical
transition the density is inversly proportional to $1/N$). Addition of a
spontaneous death process $A \stackrel{\mu}{\rightarrow} 0$ to the
model shifts the deterministic transition to $\sigma=\mu$ and
provide us with a model that admits a DP transition all the way to $N
= \infty$. In fact, our preliminary numerics show that in the Cardy-Tauber case
the scaling of the critical $\sigma$ appears to be consistent with  their prediction.
 Adding $\mu$ changes the picture
dramatically and $\sigma_c \sim \mu + N^{-\kappa}$, where $\kappa$
is identical to that measured for the SIS model. It is interesting
to note that, if the  Cardy-Tauber perturbative technique is
modified to include the spontaneous death process, at large $N$ the
distance from the DP transition is predicted to scale as $N^{-1}$ instead of
$N^{-2}$.

Many aspects of that problem are still open. In particular a
rigorous classification scheme for the quasi-deterministic behavior
is still missing. The extinction transition in higher dimensions
(where dynamic percolation admits a nontrivial critical point) and the
behavior of the system above the transition also need to be investigated. We
hope to address these subjects in future work.

\acknowledgments{We thank S. Havlin for useful remarks. The work of NMS is supported in part by the EU 6th framework CO3 pathfinder.}

\end{document}